\documentclass[12pt,preprint]{aastex}

\begin{document}

\newcommand{\sy}{{Seyfert }}
\newcommand{\etal}{{ et al. }}
\newcommand{\beq}{\begin{equation}}
\newcommand{\eeq}{\end{equation}}
\newcommand{\msun}{M_{\odot }}

\title{Inclinations and black hole masses of Seyfert 1 galaxies}
\author{Xue-Bing Wu$^{1}$ and J. L. Han$^{2}$} 
\affil{1. Department of Astronomy, Peking University, Beijing 100871, China; 
wuxb@bac.pku.edu.cn\\
	2. National Astronomical Observatories, Chinese Academy of Sciences, 
Beijing 100012, China; hjl@bao.ac.cn 
}

\begin{abstract}
A tight correlation of black hole mass (M$_{\rm{BH}}$) and central
velocity dispersion ($\sigma$) has been found recently for both active
and quiescent galaxies.  By applying this correlation, we develop a
simple method to derive the inclination angles for a sample of 11 \sy
1 galaxies that have both measured central velocity dispersions 
and black hole masses estimated by reverberation mapping. These angles,
with a mean value of 36$^o$ that agrees well with the result obtained by 
fitting the iron K$\alpha$ lines of \sy 1s observed with $ASCA$, provide 
further support to the orientation-dependent unification scheme of AGN. 
A positive correlation of the inclinations with observed FWHMs of
H$\beta$ line and a possible anti-correlation with the nuclear
radio-loudness have been found. We conclude that more accurate
knowledge on inclinations and broad line region  dynamics 
is needed to improve the black hole mass determination of AGN with the
reverberation mapping technique.
\end{abstract}

\keywords{black hole physics -- galaxies: active -- galaxies: nuclei --
                galaxies: Seyfert.}

\section{Introduction}
Supermassive black hole is believed to be an essential part of active
galaxy and quasar (Lynden-Bell 1969; Rees 1984). Recently a lot of
evidence has also been found for the existence of supermassive black 
holes in nearby quiescent galaxies (Kormendy \& Richstone 1995; Magorrian
\etal 1998). A significantly tight correlation of black hole mass
(M$_{\rm{BH}}$) with the bulge velocity dispersion ($\sigma$) was 
found for nearby galaxies (Gebhardt \etal 2000a; Ferrarese \& Merritt
2000). More recent studies indicated that some Seyfert galaxies
with M$_{\rm{BH}}$ measured by reverberation mapping method
follow the same M$_{\rm{BH}}$-$\sigma$ relation as for normal galaxies 
(Gebhardt \etal 2000b; Ferrarese \etal 2001), which implies that this
relation may be universal for galaxies. This relation  strongly suggests
a tight connection between the formation and evolution of the supermassive
black hole and the galactic bulge, though the nature of this connection
is still unclear.

The black hole masses of some active galactic nuclei (AGN) have been
recently estimated by the reverberation mapping technique (Wandel,
Peterson \& Malkan 1999; Ho 1999; Kaspi \etal 2000). With this technique,
the broad line region (BLR) size can be measured using the lag between the  
variability of continuum and emission line fluxes. The black hole mass can be
then estimated from the BLR size and the characteristic velocity (determined
by the full width at half-maximum (FWHM) of emission line). However,
it may not be so  strait-forward to estimate the characteristic velocity
of BLR according to the observed FWHMs of emission lines (Fromerth \&
Melia 2000). Many effects, especially the inclination and BLR geometry,
can lead to larger uncertainties to the estimation of black hole mass
using reverberation mapping (Krolik 2001).

Inclination is an important ingredient in the orientation dependent unified
scheme of AGN (Urry \& Padovani 1995), but it is not easy to be derived  
directly from observations. By fitting the observed iron K$\alpha$ 
line profile with the accretion disk model, Nandra \etal (1997) estimated
the inclinations, with a mean value of 30$^o$, for 18 \sy 1 galaxies.
Although with large uncertainties, their result suggested that the 
inclinations of \sy 1s are small,  consistent with Seyfert 1/2 unification
scheme (Antonucci \& Miller 1985). Moreover, numerous observations have
shown the evidence for the lack of edge-on Seyfert 1 galaxies (Keel 1980)
and suggested that \sy 1 galaxies  have the dusty torus with half opening
angles of 40$^o$ to 60$^o$ (Osterbrock \& Martel 1993; Ho, Fillippenko
 \& Sargent 1997; Schmitt \etal 2001). Therefore, the inclinations of
\sy 1 galaxies are generally expected to be small, though further
evidence is obviously needed. 

If the M$_{\rm{BH}}$-$\sigma$ relation is valid for \sy galaxies,
this provides an independent way to determine M$_{\rm{BH}}$ from the
reverberation mapping. Using  M$_{\rm{BH}}$ derived from the measured
stellar velocity dispersion and assuming a simple BLR geometry, we
develop a  method 
to derive the inclinations for
11 \sy 1 galaxies with both measured central velocity dispersions and
estimated black hole masses by reverberation mapping. Our results are
supported by several correlation studies and are consistent with
previous knowledge of inclination effects of AGN. 

\section{Black hole mass determinations of AGN}

According to the reverberation mapping technique, the black hole mass
can be estimated by a virial form,
\beq
 {\rm M_{rev}}= \frac{V^2R}{G},
\eeq
provided that the BLR region of AGN is gravitational bounded and has
a Keplerian characteristic velocity ($V$). The BLR size ($R$) can be
derived by the lag between the variability of continuum and broad
emission line. Assuming AGN having random inclinations, Wandel
\etal (1999) and Kaspi \etal (2000) related the BLR characteristic
velocity to the  FWHM of H$\beta$ emission line by 
$V=(\sqrt{3}/2)V_{\rm{FWHM}}$. However, as pointed recently by
McLure \& Dunlop (2001), the assumption of random inclinations seems
unrealistic for quasars. Many observational studies also indicated
that \sy 1 galaxies  are probably not viewed at all inclinations
(Osterbrock \& Martel 1993; Ho \etal 1997; Nandra \etal 1997;
Schmitt \etal 2001). 

On the other hand, the BLR dynamics of AGN has not been well
understood yet. The simplest case may be described as circular
orbits confined to the disk plane, but the real case is probably
much more complicated. The same as that assumed for quasars, the BLR
velocity of \sy 1s may be better represented by a combination of a
random isotropic component, with characteristic velocity
V$_r$, and a component in the disk plane, with characteristic
velocity V$_p$ (Wills \& Browne 1986). Therefore, the observed
FWHM of emission line will be given by
\beq
V_{\rm{FWHM}}=2(V_r^2+V_p^2 \sin^2 i)^{1/2},
\eeq
where $i$ is the inclination angle of the disk normal relative 
to the line of sight. Assuming that $V_p$ is significantly larger than 
$V_r$ and that $i$ lies randomly between 0 and 46$^o$, McLure \& Dunlop 
(2001) has reproduced the distribution of observed H$\beta$ FWHMs for
a sample of AGN using the above formula. If we relate the observed
emission line FWHM and BLR characteristic velocity with eq. (2), the
black hole mass of AGN can be obtained by
\beq
{\rm M_{BH}}=\frac{1}{4(\sin^2 i+{\rm A^2})} \frac{V_{\rm{FWHM}}^2 R}{G},
\eeq
where 
$A=V_r/V_p$.
In most recent studies, the black hole mass (${\rm M_{rev}}$)
determined by reverberation mapping is based on the
assumption of random inclinations and is expressed as
${\rm M_{rev}}=\frac{3}{4}\frac{V_{\rm{FWHM}}^2 R}{G}$
(Wandel \etal 1999; Kaspi \etal 2000). If the inclination effect
is considered, using eq. (3) we can derive the inclination angle by:
\beq
i= \arcsin(\sqrt{\rm \frac{M_{rev}}{3M_{BH}}-A^2}).
\eeq
 
If  ${\rm M_{BH}}$ of AGN can be derived independently by 
other methods, we can use eq. (4) to estimate the inclinations of
AGN. Fortunately a tight relation between black hole mass and central
velocity dispersion seems to exist for AGN, therefore M$_{\rm{BH}}$ can
be directly derived from the measured central velocity dispersion.
According to Gebhardt \etal (2000a), the M$_{\rm{BH}}$-$\sigma$ relation
is:
\beq
{\rm M_{BH}=1.2\times10^8 M_{\odot}} (\sigma/200 {\rm ~km~s^{-1}})^{3.75}.
\eeq
Using a slightly steeper slope  found by Ferrarese \& Merritt (2000)
has no significant effect on our results.

\tabcolsep 4.5mm
  \begin{table*}
      \caption{Data of Seyfert 1 galaxies}
         \begin{tabular}{lcccccl}
            \hline
            \noalign{\smallskip}    
Name	& $\sigma$  & Ref$^*$	  & ${\rm M_{rev}}$  		  & 	 $V_{\rm{FWHM}}(rms)$	  & log R$_{\rm nuc}$   &  Inclination  \\
	&	(km/s) &	&	($10^7M_\odot$)  &	(km/s)	&			&	($^o$)\\	
\hline	
	 \noalign{\smallskip}
3C 120	 & 	162$\pm$20  & 1 & 2.3$^{+1.5}_{-1.1}$  	 & 2210$\pm$120		 & ---	 & 22.0$^{+9.3}_{-7.7}$ \\
Mrk 79	 & 	130$\pm$9	  &  2 & 5.2$^{+2.0}_{-2.8}$  		 & 6280$\pm$850		 & ---	 & 58.5$^{+21.7}_{-27.9}$\\
Mrk 110	 & 	86$\pm$5	  &  2 & 0.56$^{+0.20}_{-0.21}$ 		 & 1670$\pm$120		 & ---	 & 37.4$^{+9.2}_{-9.5}$ \\
MrK 590		 & 169$\pm$28  & 1 & 1.78$^{+0.44}_{-0.33}$ 	 & 	2170$\pm$120	 & 	1.62	 & 17.8$^{+6.1}_{-5.9}$ \\
Mrk 817	 & 	142$\pm$6	  &  2 & 4.4$^{+1.3}_{-1.1}$  		 & 4010$\pm$180		 & 1.21	 & 41.6$^{+8.5}_{-7.5}$\\
NGC 3227	 & 144$\pm$22   & 1 &3.9$^{+2.1}_{-3.9}$   & 	5530$\pm$490	 & 	1.12	 & 37.5$^{+17.3}_{-25.4}$\\
NGC 3516	 & 124$\pm$5	 & 3 &   2.3$^{+0.69}_{-0.69}$    &       4760$\pm$240   &  		0.78  & 	38.3$\pm$7.6 \\
NGC 4051	 &  80$\pm$4	 & 2 &   0.13$^{+0.13}_{-0.08}$ 	 & 	1230$\pm$60	 & 	0.87	 & 19.6$^{+10.4}_{-6.6} $\\
NGC 4151	 & 93$\pm$5	 &   2 &1.53$^{+1.06}_{-0.89}$ 	 & 	5230$\pm$920 & 		0.49	 & 60.0$^{+30.0}_{-30.6}$ \\
NGC 4593	 & 124$\pm$29  &  1 & 0.81$^{+0.24}_{-0.24}$    	 & 3720$\pm$180     & 	---   	 & 21.6$\pm$10.5 \\
NGC 5548	 & 183$\pm$10  &  2 &12.3$^{+2.3}_{-1.8}$  		 & 5500$\pm$400		 & 1.24 & 	43.7$^{+7.6}_{-6.9}$ \\
            \hline
         \end{tabular}
\vskip 2mm
\noindent $^*$References: 1, Nelson \& Whittle (1995); 2, Ferrarese 
\etal (2001); 3, Di Nella \etal (1995).    
   \end{table*}

\section{Inclination angles of Seyfert 1 galxies}

We collected 11 \sy 1 galaxies with both measured central velocity
dispersion ($\sigma$) and estimated black hole masses (${\rm M_{rev}}$)
by reverberation mapping (Wandel \etal 1999; Ho 1999; Kaspi \etal 2000). 
Their $\sigma$ and ${\rm M_{rev}}$ values, together with the observed
FWHMs of H$\beta$ line (Wandel \etal 1999; Ho 1999) and the nuclear
radio-loudness (Ho \& Peng 2001),  are  summarized in Table 1. Seven
sources have $\sigma$ values  from high quality  measurements of
Ferrarese \etal (2001) and Di Nella \etal (1995). The $\sigma$
values of other  four sources were adopted from Nelson \& Whittle (1995). 
The uncertainties of ${\rm M_{rev}}$ and ${\rm V_{FWHM}}$ for NGC 3516
and NGC 4593  are unavailable in literature and were assumed to be
$30\%$ and $5\%$, respectively.

The inclinations of these 11 \sy 1s can be estimated using eq. (4) and
eq. (5). For simplicity we further assume ${\rm A \ll M_{rev}/3M_{BH}}$
in eq. (4), which is equivalent to the approximation that the BLR
characteristic velocity is dominated by the component in the disk
plane (McLure \& Dunlop 2001). Under this assumption, we derived the
inclinations for 11 \sy 1s (see Table 1 and Figure 1). The errors of
$i$ were estimated
from the uncertainties of both $\sigma$ and ${\rm M_{rev}}$. The 
inclination angles we derived 
are in a range from 20$^o$ to 60$^o$, with a mean value of 36.2$^o$. This
agrees with the result obtained by fitting the observed iron K$\alpha$
line in X-ray band with accretion disk model (Nandra \etal 1997), and
is consistent with the expectation of unified scheme of AGN
(Antonucci \& Miller 1985). In Figure 2, we can clearly observe an 
apparent positive correlation between inclinations and observed H$\beta$
FWHMs. A minimum $\chi^2$ fit considering the errors of both parameters 
gives, $\sin (i) = (0.23\pm0.09) +(0.086\pm0.024)
{\rm (V_{FWHM}/1000km~s^{-1})}$, with $\chi^2$ and probability of
7.13 and 62\%, respectively. The zero slope is less favored because it produces 
a larger $\chi^2$ (22.78)
with a probability of 1\% only.  A simple Spearman test gives the correlation
coefficient $R=0.78$, with have a probability of $P=4.47\times 10^{-3}$
to occur by chance. We have also applied the bootstrap method to
investigate the uncertainty of correlation coefficient by considering the 
uncertainties of both parameters, and obtained $\langle R\rangle=
0.58\pm 0.18$, which indicates a moderately significant correlation.
In Figure 3, we plotted the inclinations against the nuclear radio 
loudness, defined by the ratio between 5 GHz nuclear radio luminosity and
B-band nuclear optical luminosity (Ho \& Peng 2001). Because there are
serious contaminations to the luminosity of \sy nucleus from the host
galaxy, using the nuclear radio-loudness can maximumly diminish such
contaminations and better describe the nature of \sy nuclei. Although
only seven sources with available data for nuclear radio-loudness, 
Figure 3 still displays the trend that \sy 1s with larger nuclear
radio-loudness may have smaller inclinations. These results seem to
be  consistent with our knowledge on the inclination effects
of quasars and AGN (Wills \& Browne 1986; Wills \& Brotherton 1995).

If the characteristic velocity of BLR is dominated by the component in the
disk plane, the {\it intrinsic} FWHM of H$\beta$ emission line can be approximated by
the observed FWHM divided by $\sin(i)$. In Figure 4 we show
the distributions of intrinsic FWHMs of H$\beta$ line and M$_{\rm{BH}}$. 
Two
narrow line \sy 1s (NLS1s), NGC 4051 and Mrk 110,  locate in the lower-left
corner of Figure 4, 
though their inclination
angles, estimated to be 19.6$^o$ and 37.4$^o$, are not
significantly smaller than those of other broad line \sy 1s. This indicates
that the narrowness of emission lines of NLS1 may not be simply due to the 
inclination
effects, but is probably more related to their smaller black hole masses. These 
NLS1s may have 
accretion rates close to Eddington limit (Boller, Brandt \& Fink 1996). 
For most broad line \sy 1s, the
Eddington ratios 
($L/L_{Edd}$) are in the range of 0.1 to 0.01 (Wandel \etal 1999), which are 
significantly lower
than those of NLS1s. 

\section{Discussion}

Assuming a Kelperian rotation of BLR and  \sy 1s following the same 
M$_{\rm{BH}}$-$\sigma$ relation as normal galaxies, we develop a
method to derive the inclinations of 11 well studied \sy 1 galaxies. 
The values of  these inclinations derived by us  seem to be  
supported by a positive correlation with the observed H$\beta$
FWHMs and a possible anti-correlation with the nuclear radio loudness.
Our results agree well  with previous knowledge on inclinations
of \sy 1s and are consistent with the expectation of unification
scheme of AGN.

It is worth noting that the BLR dynamics may not be as simple as
we assumed, though the Keplerian rotation of BLR have been confirmed
by some recent studies (Peterson \& Wandel 1999, 2000). The inclinations
we derived for 11 \sy 1s may be regarded as upper limits because we
assumed the characteristic velocity in the disk plane is the dominant
component in BLR. Considering of the isotropic component will reduce
the values of inclinations (see eq. (4)).  VLBI study on the radio-loud
\sy 1 galaxy 3C~120 derived a lower inclination angle (about 10$^o$)
based on the synchrotron self-Compton jet model (Ghisellini \etal 1993),
which is smaller than but still marginally consistent with the value 
obtained by us.

We noticed that ratio between the black hole mass determined by 
reverberation mapping (M$_{\rm rev}$) and the `true' black hole mass
(M$_{\rm BH}$), can be approximated by $3\sin^2(i)$ (see eq. (4)).
With the mean value of $i$ (36.2$^o$) derived by us, we obtained
M$_{\rm rev}$/M$_{\rm BH}=1.05$ for \sy 1s.  It means that the black 
hole mass estimated by the standard method of reverberation mapping,
can still represent the `true' black hole mass well if the inclination
of \sy galaxy is not substantially different from 36.2$^o$. This may
be the reason why \sy galaxies also follow the M$_{\rm{BH}}$-$\sigma$
relation  even if  the values of M$_{\rm{BH}}$  measured by reverberation
mapping were adopted (Gebhardt \etal 2000b; Ferrarese \etal 2001).
However, the M$_{\rm rev}$/M$_{\rm BH}$ value changes from 0.35 to
2.25 if the inclination increases from 20$^o$ to 60$^o$. Therefore, 
more accurate information about inclinations and the BLR dynamics of AGN
will be undoubtedly helpful to the improvement of central black hole
mass estimations with the reverberation mapping  technique.

\acknowledgments

We thank the anonymous referee for helpful suggestions and Xiaohui
Sun for assistance on statistics.  This work was supported by
the Pandeng Project, the National Natural Science Foundation, and the
National Key Basic Research Science Foundation (NKBRSF G19990752) 
in China.

\newpage

\begin{figure}
\plotone{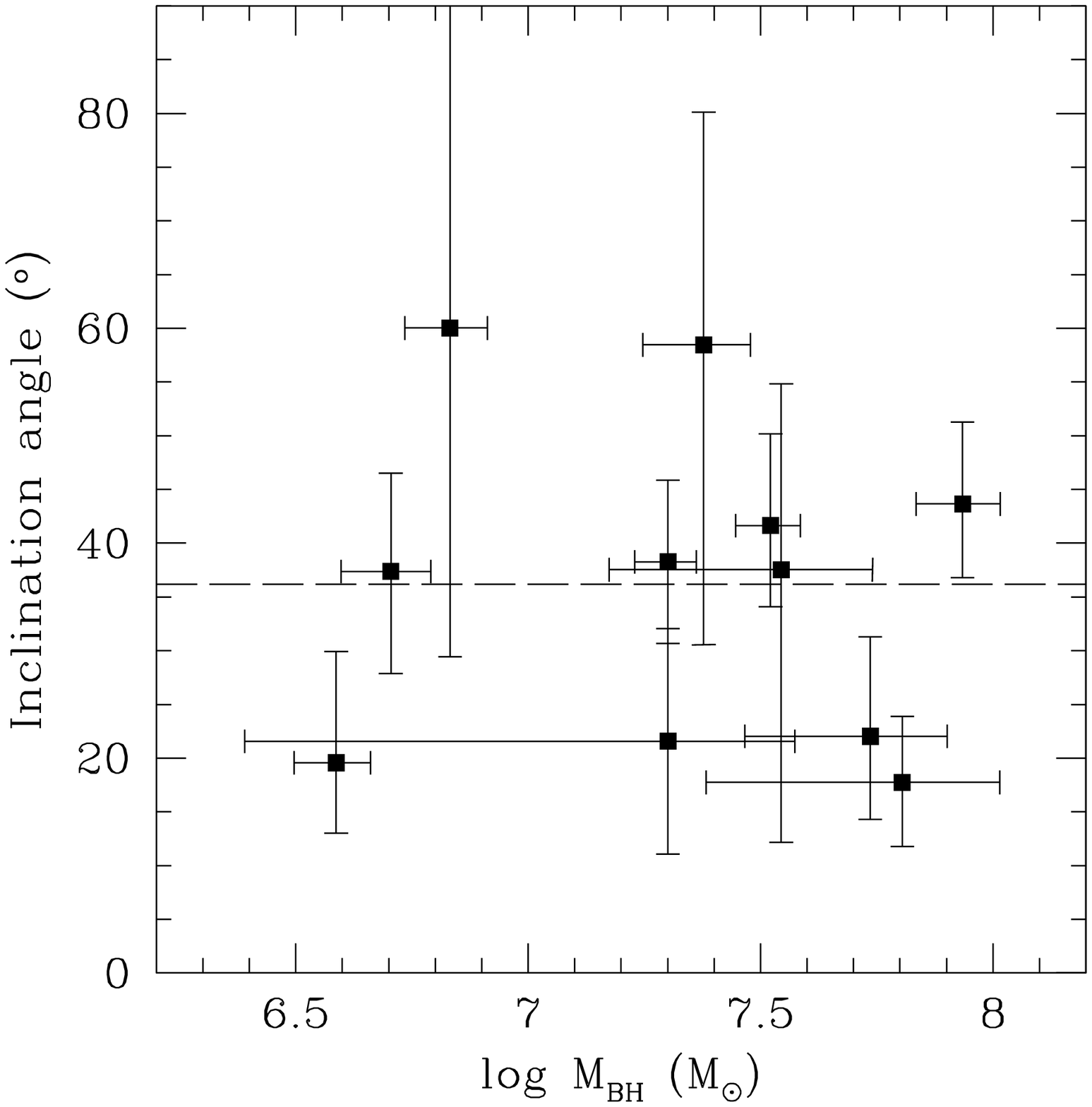}
\caption{The derived inclination angles of 11 \sy 1 galaxies
against the black hole masses 
determined by the M$_{\rm{BH}}$-$\sigma$ relation. 
The dashed line represents the mean value
$\langle i \rangle =36.2^o$.}
\end{figure}

\begin{figure}
\plotone{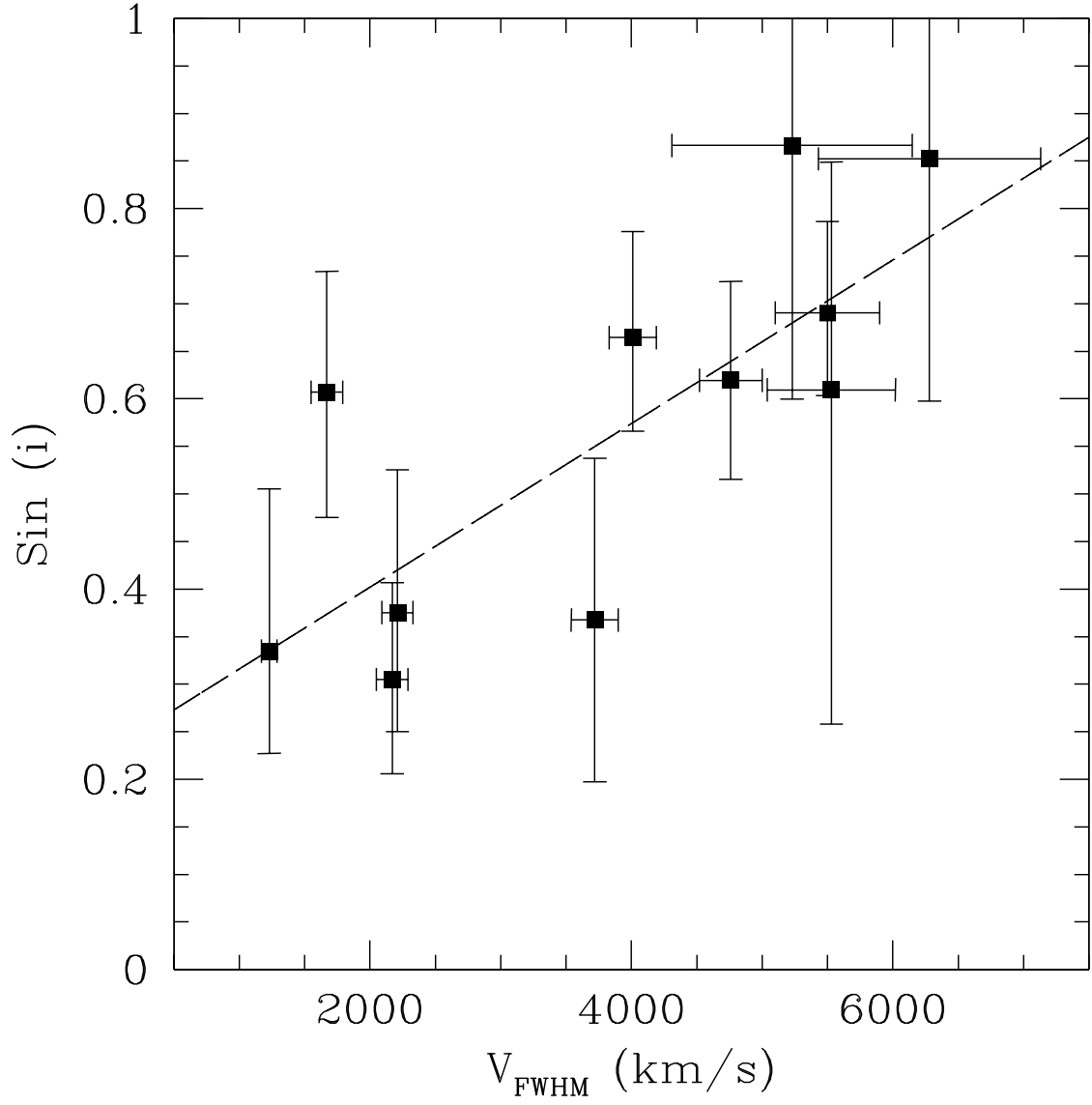}
\caption{The inclination angles against the observed FWHMs of H$\beta$ line. 
The dashed line shows the minimum $\chi^2$ fit with errors of both
parameters considered.
}
\end{figure}

\begin{figure}
\plotone{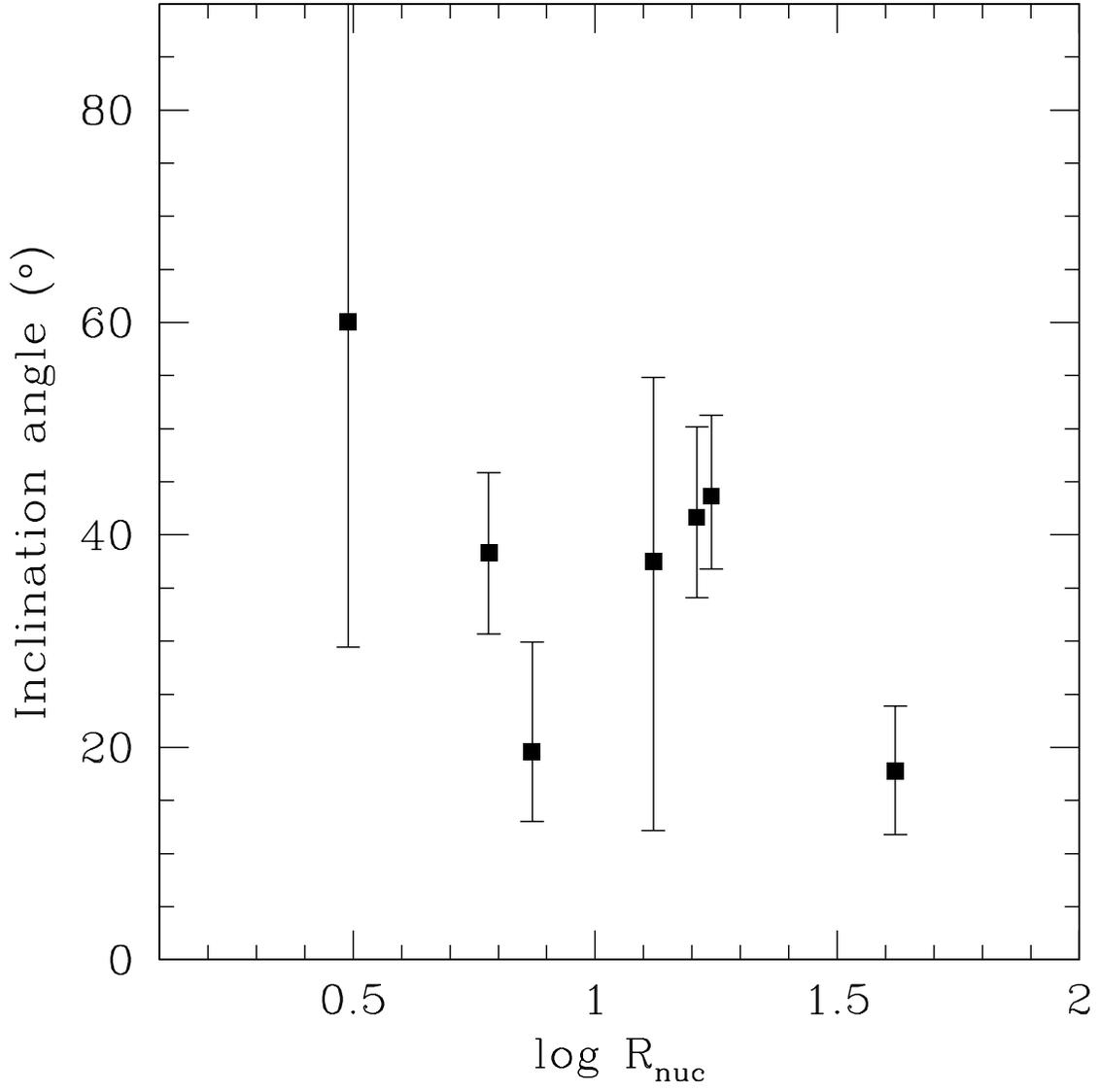}
\caption{The inclination angles against the nuclear radio-loudness.}
\end{figure}

\begin{figure}
\plotone{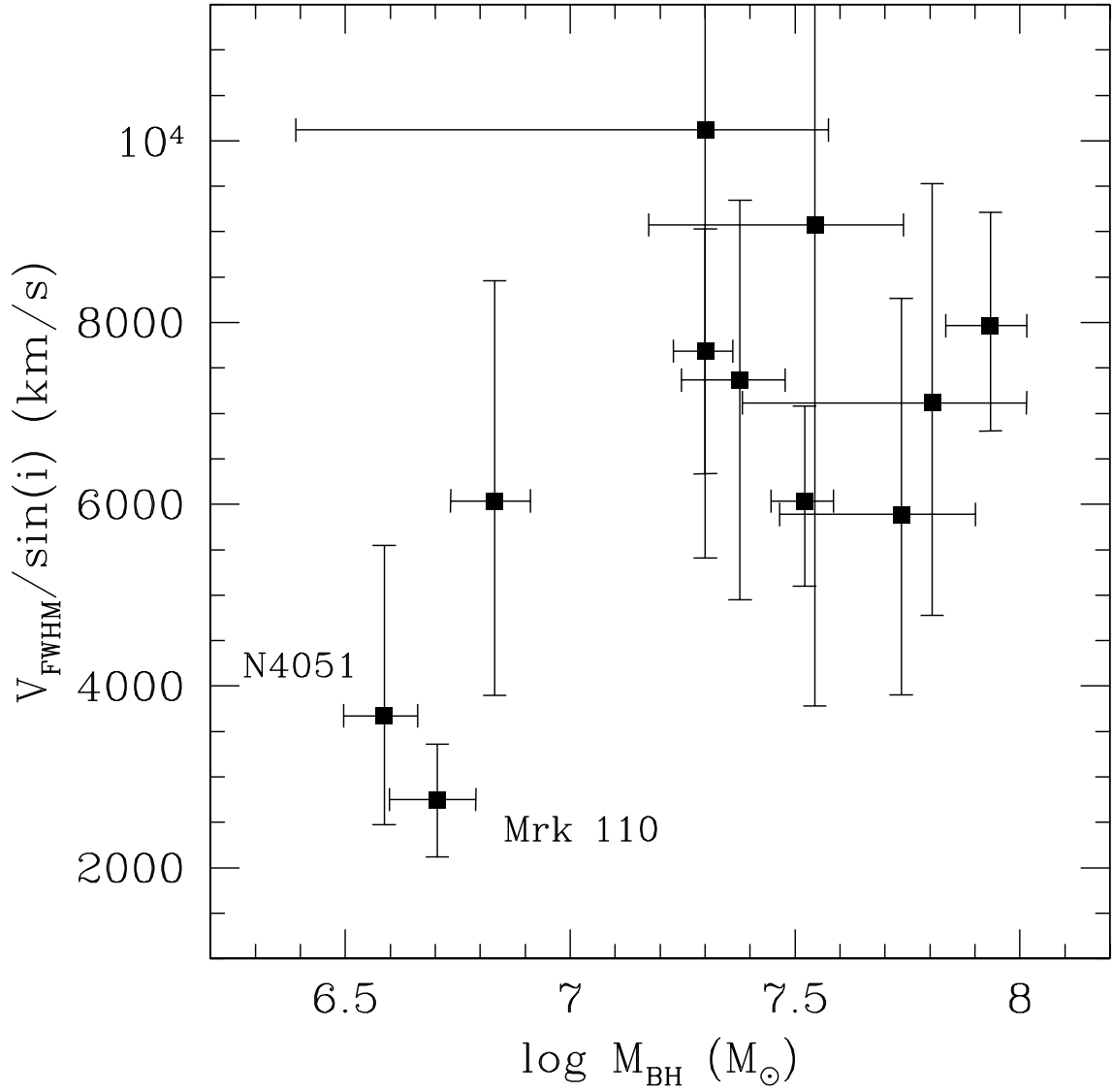}
\caption{The intrinsic FWHMs of H$\beta$ line against black hole masses determined
by the M$_{\rm{BH}}$-$\sigma$ relation.}
\end{figure}


\begin{thebibliography}{}

\bibitem[1985]{antonucci85} Antonucci, R.R.J., Miller, J.S. 1985, ApJ, 297, 621
\bibitem[1996]{boller96} Boller, Th., Brandt, W.N., Fink, F. 1996, A\&A, 305, 53
\bibitem[1995]{dinella95} Di Nella, H., Farcia, A.M., Faruier, R., Paturel, G.
1995, A\&ASS, 113, 151
\bibitem[2000]{ferrarese00} Ferrarese, L., \& Merritt, D. 2000, ApJ, 539, L9
\bibitem[2001]{ferrarese01} Ferrarese, L., \etal 2001, ApJ, 555, L79
\bibitem[2000]{fromerth00} Fromerth, M.J., \& Melia, F. 2000, ApJ, 533, 172
\bibitem[2000]{gebhardt00a} Gebhardt, K. \etal 2000a, ApJ, 539, L13
\bibitem[2000]{gebhardt00b} Gebhardt, K. \etal 2000b, ApJ, 543, L5
\bibitem[1993]{ghisellini93} Ghisellini, G., Padovani, P., Celotti, A., Maraschi, 
L. 1993, ApJ, 407, 65
\bibitem[1999]{ho99} Ho, L.C. 1999, in Observational Evidence for Black Holes in the Universe,
	ed. S.K. Charkrabarti (Dordrecht: Kluwer), 157
\bibitem[1997]{ho97a}  Ho, L.C., Filippenko, A.V., \& Sargent, W.L.W. 1997, ApJS, 112, 
	315
\bibitem[2001]{ho01a} Ho, L.C. \& Peng, C.Y. 2001, ApJ, 555, 650
\bibitem[2000]{kaspi00} Kaspi, S., Smith, P.S., Netzer, H., Maoz, D., Jannuzi, B.T., 
	\& Giveon, U. 2000, ApJ, 533, 631
  \bibitem[1980]{keel80} Keel, W.C. 1980, AJ, 85, 198
\bibitem[1995]{kormendy} Komendy, J. \& Richstone, D. 1995, ARA\&A, 33 581
\bibitem[2001]{krolik01} Krolik, J.H. 2001, ApJ, 551, 72 
  \bibitem[1969]{lyden-bell} Lyden-Bell, D. 1969, Nature, 223,690 
      \bibitem[1998]{magorrian} Magorrian, J., \etal 1998, AJ, 115, 2285
\bibitem[2001]{mclure} McLure, R.J,  Dunlop, J.S. 2001, MNRAS, in press (astro-ph/0009406)
\bibitem[1997]{nandra97} Nandra, K., George, I.M., Mushotzky, R.F., Turner, T.J.,
Yaqoob, T. 1997, ApJ, 477, 602
\bibitem[1995]{nelson95} Nelson, C.H., \& Whittle, M. 1995, ApJS, 99, 67
\bibitem[1993]{osterbrock} Osterbrock, D.E. \& Martel, A. 1993, ApJ, 414, 552
\bibitem[1999]{peterson99} Peterson, B. \& Wandel, A. 1999, ApJ, 521, L95
\bibitem[1999]{peterson00} Peterson, B. \& Wandel, A. 2000, ApJ, 540, L13
   \bibitem[1984]{rees} Rees, M.J. 1984, ARA\&A, 22 471
\bibitem[2001]{schmitt} Schmitt, H.R., Antonucci, R.R.J., Ulvestad, J.S., Kinney, A.L.,
Clarke, C.J., Pringle, J.E. 2001, ApJ, 555, 663
\bibitem[1995]{urry95} Urry, C.M. \& Padovani, P. 1995, PASP, 107, 803
\bibitem[1999]{wandel99b} Wandel, A., Peterson, B.M., \& Malkan, M.A.  1999, ApJ, 526, 579
\bibitem[1986]{wills86} Wills, B.J., Browne, I.W.A. 1986, ApJ, 302, 56
\bibitem[1995]{wills95} Wills, B.J., Botherton, M.S. 1995, ApJ, 448, L81

\end{thebibliography}
\end{document}